\documentstyle[latexsym,amssymb,epsfig,aps,floats,preprint,amsfonts]{revtex}

\def\laq{\raise 0.4ex\hbox{$<$}\kern -0.8em\lower 0.62ex\hbox{$\sim$}}
\def\gaq{\raise 0.4ex\hbox{$>$}\kern -0.7em\lower 0.62ex\hbox{$\sim$}}

\font\tenbb=msbm10
\font\sevenbb=msbm7
\font\fivebb=msbm5
\newfam\bbfam
\textfont\bbfam=\tenbb \scriptfont\bbfam=\sevenbb
\scriptscriptfont\bbfam=\fivebb

\newcommand{\beq}{\begin{equation}} 
\newcommand{\eeq}{\end{equation}}
\newcommand{\bea}{\begin{eqnarray}} 
\newcommand{\eea}{\end{eqnarray}}
\newcommand{\ba}{\begin{array}} 
\newcommand{\ea}{\end{array}}

\begin{document}
\draft
\preprint{\vbox{\baselineskip=12pt
\rightline{GRP/00/549} 
\vskip 0.2truecm
}}

\title{Optical noise correlations and beating the standard
quantum limit in advanced gravitational-wave detectors}
\author{Alessandra Buonanno and Yanbei Chen}
\address{Theoretical Astrophysics and Relativity Group\\
California Institute of Technology, Pasadena, California 91125, USA}
\maketitle
\begin{abstract}
The uncertainty principle, applied naively to the test masses of a 
laser-interferometer gravitational-wave detector, produces a {\it
standard quantum limit} (SQL) on the interferometer's sensitivity.
It has long been thought that beating this SQL would require
a radical redesign of interferometers.  However, we show that 
LIGO-II interferometers, currently planned for 2006, can beat the SQL by 
as much as a factor two over a bandwidth $\Delta f \! \sim \! f$,
if their thermal noise can be pushed low enough.  This is due to
dynamical correlations between photon shot noise and radiation-pressure 
noise, produced by the LIGO-II signal-recycling mirror.
\end{abstract}

\begin{figure}
\begin{center}
\epsfig{file=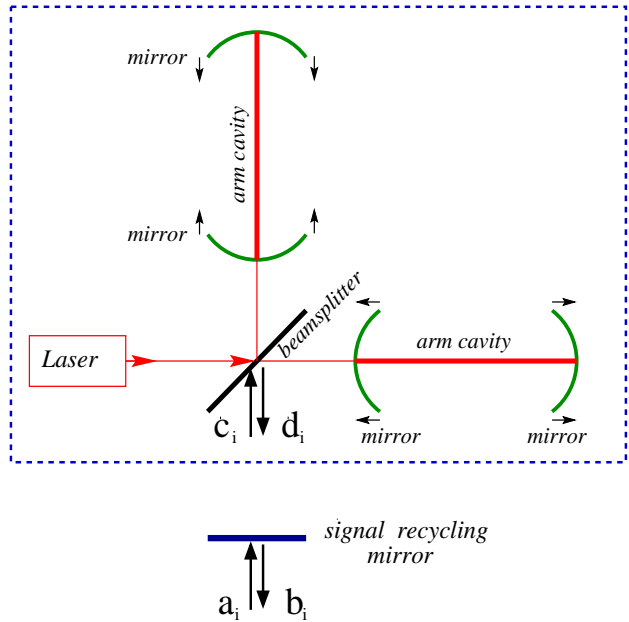,width=0.5\textwidth,height = 0.5\textwidth,angle=-90}
\vskip 0.2truecm
\caption{ Schematic view of a {\rm LIGO-II} signal recycled (SR)
interferometer. The interior of the dashed box is a conventional {\rm
LIGO-I} interferometer; $c_i$ and $d_i$ 
are the input and output fields
at the beam splitter's dark port; $a_i$ and $b_i$ are the full system's
vacuum input 
and signal output. The arrows indicate 
gravity-wave-induced mirror displacements.}
\label{Fig1}
\end{center}
\end{figure}

\begin{figure}
\centerline{
\epsfig{file=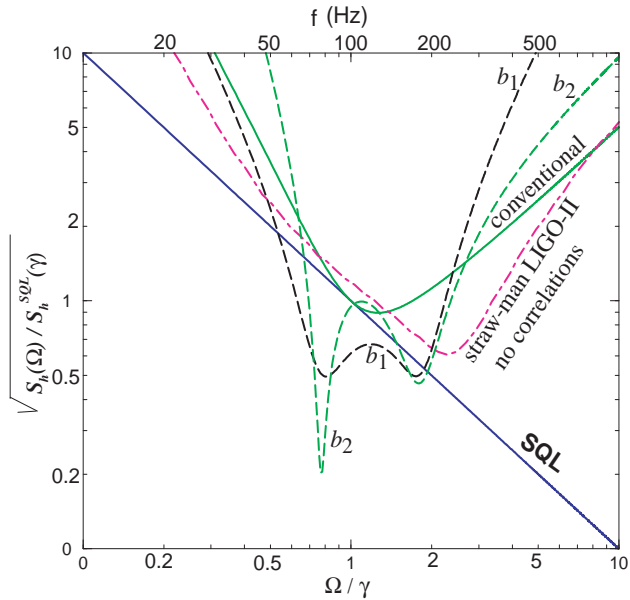,width=0.5\textwidth,height = 0.5\textwidth,angle=-90}}
\vskip 0.1truecm
\caption
{Log-log plot of $\sqrt{S_h(\Omega)/S_h^{\rm SQL}(\gamma)}$ 
versus $\Omega/\gamma$ for the quadratures $b_1$ 
($\zeta = \pi/2$) and $b_2$ ($\zeta = 0$) with
$\rho = 0.9$, $\phi = \pi/2 -0.47$ and $I_0 = I_{\rm SQL}$,
for the SQL, for a conventional interferometer with $I_0=I_{\rm SQL}$,
and for a straw-man LIGO-II design [2] with shot-noise / radiation-pressure
correlations naively omitted.
For LIGO-II, 
$\gamma =2\pi \times 100\,{\rm Hz}$ (top axis) and 
$\sqrt{S_h^{\rm SQL}(\gamma)} = 2\times 10^{-24}\,{\rm Hz}^{-1/2}$. 
These curves do not include 
seismic and thermal noises; for LIGO-II the latter is currently estimated to be 
slightly above the SQL [8].} 
\label{Fig2}
\end{figure}

\newpage
A laser-interferometer gravitational-wave detector (``interferometer'' 
for short) consists mainly of an L-shaped assemblage
of four mirror-endowed test masses, suspended from seismic-isolation 
stacks (see Fig.~\ref{Fig1}). Laser interferometry is used to monitor
changes in the 
relative positions of the test masses produced by
gravitational waves.  
The uncertainty principle states that, if the relative positions 
are measured with high precision, then the test-mass momenta 
will thereby be perturbed.  As time passes, the momentum perturbations 
will produce position uncertainties, thereby possibly masking
the tiny displacements produced by gravitational waves.  A detailed
analysis of this process gives rise to the standard quantum limit (SQL)
for interferometers: a limiting (single-sided)
noise spectral density $S_h^{\rm SQL} = 8 \hbar/(m\Omega^2L^2)$ for the 
dimensionless gravitational-wave (GW) signal $h(t) = \Delta L/L$ \cite{KT80}.  
Here $m$ is the mass of each identical 
test mass, $L$ is the length of the interferometer's arms, 
$\Delta L$ is the time evolving difference in the 
arm lengths, $\Omega$ is the GW angular frequency, and $\hbar$ is Planck's constant.  
This SQL is shown in Fig.~\ref{Fig2} 
for the parameters of LIGO-II \cite{SR} [the second generation interferometers 
in Laser Interferometer Gravitational Observatory 
(LIGO), planned to operate in $\sim 2007$--2009]: $m=30$ kg, $L= 4$ km.  
The ``straw-man'' design for LIGO-II \cite{GSSW99}, 
assuming (naively) no correlations between photon 
shot noise and radiation-pressure noise,
is capable of going very close and parallel to 
the SQL over a wide frequency band: $\sim 50$ Hz to $\sim 200$ Hz 
(see Fig.~\ref{Fig2}).  

Braginsky, who formulated the concept of SQL's for high-precision
measurements~\cite{B68-70s}, also demonstrated that it is possible to 
circumvent SQL's by changing the designs of one's
instruments \cite{B68-70s}, \cite{BK92}.  
Since the 1970s, it has been thought that for GW interferometers 
the redesign must be major ---
e.g., injecting squeezed vacuum into an interferometer's 
dark port \cite{SFD} and/or introducing 4km-long filter cavities into 
the interferometer's output port, as has recently been proposed for
LIGO-III \cite{KLMTV00} to implement frequency-dependent 
homodyne detection \cite{HFD}. 
Yuen and Ozawa have also conceived ways to beat the SQL by taking 
advantage of the so-called contractive states \cite{debate}, 
but it is not yet clear how to implement their ideas in real
GW interferometers.

In this paper we show that although major redesigns could not 
ba avoided if we want to beat the SQL significantly,  
however LIGO-II interferometers with their currently planned design, 
can beat the SQL by modest amounts (see, e.g., noise curves 
$b_1$ and $b_2$ in Fig.~\ref{Fig2}), {\it if} 
all sources of thermal noise can also be pushed below the SQL.  
For current LIGO-II designs, estimates place 
the  dominant, thermoelastic component at about the SQL \cite{BGV00}. 

As is well known, there are two aspects of the uncertainty principle:
(i) the quantum mechanics of the test-mass wave function, and (ii)
the Heisenberg-microscope-like influence of the laser light used to
measure the position. Braginsky and colleagues 
\cite{BK92,BGKMTV00} have shown that 
{\it the test-mass wave-function aspect of the uncertainty principle 
is irrelevant to the operation of a GW interferometer}.  
Indeed, the interferometer does {\it not} measure 
relative test-mass positions; it only monitors 
classical-force-induced changes in the relative positions, and those 
changes, in the LIGO frequency band, are not contaminated 
by the details of the test-mass wave functions.  
As a result, {\it the light is the only enforcer of the SQL}.

Braginsky and Khalili have also shown that \cite{BK92} 
as long as there are no correlations between 
the light's shot noise and its radiation-pressure-fluctuation noise, 
the light firmly enforces the SQL. This is the case 
for ``conventional interferometers'', i.e. for interferometers that
have no signal-recycling mirror on the output port and a simple
(frequency independent) homodyne detection is performed (the type of
interferometer used, e.g., in LIGO-I/Virgo).  However, 
the signal-recycling mirror \cite{SR} (which is being 
planned for LIGO-II as a tool to reshape the noise curves 
\footnote{A power-recycling mirror
is also used in real interferometers to increase the light
power at the beamsplitter, but it will not affect the quantum noise in
the dark-port output. For this reason we do not take it into account.}), 
sends back into the arm cavities the signal 
coming out from the dark port and thereby 
produces shot-noise / 
back-action-noise correlations, which 
{\it break the light's ability to enforce the SQL}.
These {\it dynamical} correlations arise naturally from the 
nontrivial coupling between the antisymmetric mode of motion  of 
the four arm-cavity mirrors and the signal recycled 
optical fields \cite{BC200}. This coupling invalidates  
the naive picture, according to which the arm cavity mirrors 
behave like free test masses 
subject only to Poissonian quantum-vacuum fluctuations. 
As we show below, the interferometer as a whole 
responds to a GW signal as an ``optical 
spring'' and this oscillatory behaviour is responsible
for the resonant amplification of the GW signal and 
the beating of the SQL. Braginsky, Gorodetsky and Khalili 
\cite{OB}, in designing the ``optical bar'' GW detectors, 
were the first to suggest that this phenomenon 
could be used to reach sensitivities beyond the 
free-mass SQL. The resonant dips in Fig.~\ref{Fig2} correspond to the 
resonant frequencies of the two dimensional dynamical system 
formed by the arm-cavity mirrors and the signal-recycled optical field. 
Hence, the SR interferometer's dynamics 
cannot be described by a successive sequence of independent 
measurements of the test-mass displacements, for which 
the SQL was originally derived \cite{B68-70s} 
and further discussed \cite{debate}. 
On the contrary, the SQL for a free test mass is no longer 
relevant for SR interferometers. Its only remaining role is as a 
reminder of the regime where back-action noise is comparable to the 
shot noise. 
The remainder of this paper is devoted to explaining these claims.
The full details will be published elsewhere \cite{BC200,BC300}.

Kimble, Levin, Matsko, Thorne and Vyatchanin have recently derived the
input-output relations for a 
conventional interferometer~\cite{KLMTV00} 
using the Caves-Schumaker two-photon formalism~\cite{CS85}. 
The full electric field, in the Heisenberg picture, 
at the output (dark) port, i.e. soon after
the beamsplitter (see Fig.~\ref{Fig1}), reads:
\bea
\label{eq1}
E(t) &=& \sqrt{\frac{4 \pi \hbar \omega_0}{{\cal A} c}}
\left [ \cos (\omega_0 t) \int_0^\infty 
({d}_1 e^{-i\Omega t} + {d}_1^\dagger e^{i\Omega t})
\frac{d \Omega}{2 \pi} \right . \nonumber \\
&& + \left . \sin (\omega_0 t) \int_0^\infty 
({d}_2 e^{-i\Omega t} + {d}_2^\dagger e^{i\Omega t})
\frac{d \Omega}{2 \pi} \right ]\,,
\eea
where ${d}_1$ and ${d}_2$ are the two output quadratures (see Fig.~\ref{Fig1}), 
$\omega_0$ is the carrier angular frequency, ${\cal A}$ is the 
effective cross sectional area of the laser beam and $c$ is the speed of 
light. 
Indicating by ${c}_1$ and ${c}_2$ the two input quadratures 
at the dark port, the input-output relations, at side-band (gravity-wave) 
angular frequency $\Omega$, are~\cite{KLMTV00}: 
\beq
\label{eq2}
{d}_1 = {c}_1\, e^{2 i \beta}\,, \quad 
{d}_2 = ({c}_2-{\cal K} {c}_1)\, 
e^{2 i \beta}+\frac{h\sqrt{2{\cal K}} e^{i \beta}}{h_{\rm SQL}}\,,
\eeq
where $2\beta=2\arctan{{\Omega}/{\gamma}}$ is the 
net phase gained by the field at sideband frequency $\Omega$ while in
the arm cavity, $\gamma = Tc/4L$ is the 
half bandwidth of the arm cavity ($T$ is the power 
transmissivity of the input mirrors); 
$h$ is the Fourier transform of the 
GW field, and $h_{\rm SQL} \equiv \sqrt{S_h^{\rm SQL}}$
is the SQL for GW detection. 
The quantity ${\cal K}={2 (I_0/I_{\rm SQL}) 
\gamma^4}/{(\Omega^2(\gamma^2+\Omega^2))}$ in Eq.~(\ref{eq1}) 
is the effective coupling constant which relates 
the motion of the test mass to the output signal.
Finally, $I_0$ is the input light power at the beamsplitter, while $I_{\rm SQL} = 
{m L^2 \gamma^4}/{(4 \omega_0)}$ is the power 
needed by a conventional interferometer to reach the SQL.
We indicate by $l$ the length of the SR cavity 
and limit our analysis to a SR cavity much shorter than 
the arm cavities, e.g., $l \simeq 10\,{\rm m}$. 
We introduce $\phi\equiv [\omega_0 l/c]_{{\rm mod}\, 2\pi} $, 
the phase gained by the carrier while traveling one way in the SR cavity. 

Propagating the electric field (\ref{eq1}) down to the SR 
mirror and introducing the input and output quadratures
$a_i$ and $b_i$ ($i = 1,2$) for the entire 
SR interferometer
(Fig.~\ref{Fig1}), we obtain the final input-output relations \cite{BC200}:
\footnote{
Here we face a delicacy of the Fourier-based formalism 
due to possible unstable modes of the system. 
We cured the problem by introducing an appropriate control system 
which leaves the expression of the noise spectral density 
unchanged \cite{BC200,BC300}.}
\beq
\left (\matrix {b_1 \cr b_2} \right)=
\frac{1}{M}\left[
e^{2 \,i \beta}
\left(\matrix {C_{11}&C_{12}\cr C_{21}&C_{22}}\right)
\left(\matrix{ a_1 \cr  a_2}\right) 
+ 
\sqrt{2 {\cal K}}\tau\,e^{i\beta}
\left(\matrix{ D_1 \cr D_2}\right)
\frac{h}{h_{\rm SQL}}
\right],
\label{eq3}
\eeq
where, to ease the notation, we have defined:
\bea
&& M=1 + \rho^2\, e^{4\,i\beta}-
2\rho \, e^{2\,i\beta}\left(\cos{2\phi}+\frac{{\cal K}}{2}\,
\sin{2\phi} \right), \nonumber \\
&& C_{11}=C_{22}=(1+\rho^2)
\left(\cos{2\phi}+\frac{{\cal K}}{2}\sin{2\phi} \right) 
-2\rho\cos{2\beta}\,, \nonumber \\
&& C_{12}=-\tau^2 \left(\sin{2\phi}+{\cal K}\,\sin^2{\phi}\right), 
\quad \quad C_{21}=+ \tau^2\left(\sin{2\phi}-{\cal K}\,\cos^2{\phi}\right), \nonumber \\
&& D_{1} = - \left(1 +\rho\, e^{2\,i\beta} \right)\,\sin{\phi}\,,
\quad \quad D_{2} = - \left(-1 +\rho\, e^{2\,i\beta} \right)\,\cos{\phi}\,.
\label{eq4}
\eea
In the above equations $\rho$ and $\tau$ are the amplitude 
reflectivity and transmissivity of the SR mirror, respectively. 
For a lossless SR mirror: 
$\tau^2 + \rho^2 = 1$. Because $a_i, a_i^\dagger$ in Eq.~(\ref{eq3}) 
represent a free field, they satisfy the usual commutation relations for 
quadratures with $\Omega \ll \omega_0$\cite{CS85}.

We assume a frequency-independent quadrature 
$b_\zeta = b_1 \sin \zeta + b_2 \cos \zeta$ 
is measured via homodyne detection, and the noise is calculated as 
follows \cite{KLMTV00}.
We define $h_n(\Omega) \equiv \Delta 
b_{\zeta}{h_{\rm SQL}M}/[{\sqrt{2{\cal K}}\tau}
(D_1\sin \zeta + D_2\cos \zeta)]$,
where $\Delta b_{\zeta}$ is the noise part of $b_\zeta$, and  
then the (single-sided) spectral density $S_h(f)$
of $h_n$, with $f = \Omega/2\pi$, can be computed by the formula:
$2\pi \delta(\Omega - \Omega^\prime) S_h(f) =
\langle  h_n(\Omega) h_n^\dagger(\Omega^\prime) + 
h_n^\dagger(\Omega^\prime) h_n(\Omega) \rangle$.
Assuming that the input is in its vacuum state,   
we find \cite{BC200} that the noise spectral density can 
be written in the simple form (note that $C_{ij}\in\Re$):
\bea
\label{eq5}
&& S_h = \frac{h_{\rm SQL}^2}{2{\cal K}}\,
\frac{1}{\tau^2\,\left|D_1\,\sin\zeta+D_2\,\cos\zeta\right|^2}\times \nonumber \\
&&\left [ \left(C_{11} \sin\zeta+C_{21} \cos\zeta\right)^2 + 
\left(C_{12} \sin\zeta+C_{22} \cos\zeta\right)^2 \right ].
\eea

Fig.~\ref{Fig2} shows this $S_h(f)$
for the two quadratures 
$b_1$ (i.e.\ $\zeta=\pi/2$) and $b_2$ ($\zeta=0$), with (for definiteness)
$\rho = 0.9$, $\phi = \pi/2 -0.47$ and $I_0 = I_{\rm SQL}$.  Also shown 
for comparison are the SQL, and $S_h(f)$ for a straw-man LIGO-II design
when the shot-noise/radiation-pressure
correlations are (naively) ignored \cite{GSSW99}, and for a 
conventional interferometer with $I_0=I_{\rm SQL}$. 
The sensitivity curves for the two quadratures 
go substantially below the SQL and 
show an interesting resonance structure.
To explain the resonant frequencies
in the case of a highly-reflecting  
SR mirror, we have found it convenient to investigate the {\it free} 
oscillation modes of the entire interferometer. By {\it free} we mean 
no GW signals [$h(\Omega)=0$] and perfect reflectivity for the SR mirror 
($\rho=1$). The free-oscillation frequencies satisfy the relation
\cite{BC200,BC300}: 
$\cos 2\beta = \cos 2 \phi+ {\cal K} \sin \phi \cos \phi$,
which can be solved to give
$ {\Omega^2_{\rm res}}/{\gamma^2} = [ \tan^2 \phi \pm 
\sqrt{\tan^4 \phi - {4I_0}/{I_{\rm SQL}}\,\tan \phi}]/2$,  
which agrees quite accurately with the frequencies of the valleys in the
dashed noise curves ($\rho \,\laq\,1$) of Fig.~\ref{Fig2}. 
For very low light power ($I_o \ll I_{\rm SQL}$) the resonant 
frequencies decouple into: $\Omega_{\rm res}^0 \simeq 0$, i.e. the 
eigenfrequency of a free mass 
and  $\Omega_{\rm res}^{1,2} \simeq \pm \gamma \tan\phi$, i.e. the 
optical resonances of a SR interferometer with fixed 
arm-cavity mirrors \cite{SR}.
By increasing the light power up to $I_o =I_{\rm SQL}$,
the test masses and the optical field get more and more coupled, 
and the resonant frequencies of the entire system  
become a ``mixture'' of the two decoupled resonances. 
It is easy to show \cite{BC200,BC300} that 
the low-frequency resonant dip in Fig.~\ref{Fig2} originates 
from the free-mass eigenfrequency $\Omega_{\rm res}^0$, 
modified by the dependence of the radiation-pressure force  on 
the test-mass motion's history; while the higher-frequency 
resonant valley is largely due to the optical field 
resonances $\Omega_{\rm res}^{1,2}$. Hence, the SR mirror 
feeds back the signal into the arm cavities and makes 
the SR interferometer behave as an ``optical spring'' detector. 
The GW device gains sensitivity near the resonant frequencies. 

To give a first rough idea of the performances that 
a SR interferometer 
with homodyne detection can reach
{\it if thermal noise can be made negligible}, 
we have estimated the signal-to-noise ratio 
$(S/N)^2 = 4\int_{0}^{\infty}
{|h(f)|^2}/{S_h(f)} df$ \cite{KT80} 
for gravitational waves from 
binary systems made of black holes and/or neutron stars.  Using
the Newtonian, quadrupole approximation for which
the waveform's Fourier transform is
$|h(f)|^2 \propto f^{-7/3}$, and 
introducing in the above integral a lower cutoff due to seismic  
noise at $\Omega_s = 0.1\gamma$ ($f_s \simeq 10\,{\rm Hz}$), 
we get for the parameters used in Fig.~\ref{Fig2}: 
${({S}/{N})_{1}}/{({S}/{N})_{\rm conv.}} \simeq 1.83$ and 
${({S}/{N})_{2}}/{({S}/{N})_{\rm conv.}} \simeq 1.98$.
These numbers refer to the first and second 
quadratures, respectively. 
Here ${({S}/{N})_{\rm conv.}}$ is the signal to noise ratio
given by a conventional interferometer with the same light-power input
at the beamsplitter, which is SQL-bounded.

We now briefly discuss how optical losses affect 
the noise in a SR interferometer.  We have computed~\cite{BC200} 
the influence of losses using (i) 
the lossy input-output relations [analog of Eq.\ (\ref{eq2})] for 
a conventional interferometer [boxed part of Fig.\ \ref{Fig1}] as derived
in~\cite{KLMTV00}, and (ii) an analogous treatment of losses in the SR
cavity.  We find that for loss levels expected in LIGO-II~\cite{GSSW99}, 
the optical losses have only a moderate influence on the noise curves 
of Fig.~\ref{Fig2}; primarily, they just smooth out the deep resonant valleys. 
More specifically, for (i) the physical parameters used in Fig.~\ref{Fig2}, 
(ii) a net fractional photon loss of 1\% in the arm cavities and 2\% in each
round trip in the SR cavity, and (iii) a photodetector efficiency of 90\%,
we find a fractional loss in $S/N$ for inspiraling binaries of 
$8\%$ and $21\%$, for the first and second 
quadratures, respectively.

In the last part of this letter we discuss the role 
played by the shot-noise / radiation-pressure correlations, 
present in LIGO-II's noise spectral density (\ref{eq5}), 
in beating the SQL. Our analysis is based on the general formulation 
of linear quantum measurement theory developed by Braginsky and Khalili 
in~\cite{BK92}. 
Quite generically 
\cite{BK92,BC300}, we can rewrite the output ${\cal O}$ of the whole optical system as:
$ {\cal O} = {\cal Z}+ {\cal R}_{xx}\,{\cal F} + L\,h$.
Here by output we mean one of the two quadratures
$b_1$, $b_2$ or a combination of them, 
e.g., $b_\zeta$ (modulo a normalization factor). ${\cal R}_{xx}$ 
in the above equation is the 
susceptibility of the antisymmetric mode 
of motion of the four mirrors~\cite{BK92}, given by
${\cal R}_{xx}(\Omega)=-{4}/{(m\Omega^2)}$;  
${\cal Z}$  is the {\it effective}\footnote{
We refer to ${\cal Z}$ and 
${\cal F}$ as effective because we have shown \cite{BC300} that 
for a SR interferometer the {\it real} force  
acting on the test masses is a combination of these effective fields. 
When the shot noise and radiation-pressure-noise are correlated, the real 
force does not commute with itself at different times~\cite{BC300},
which makes the analysis in terms of real quantities more complicated
than in terms of the effective ones.}
 output noise field 
and ${\cal F}$ is the {\it effective} back-action 
force, both of these operators do not depend on the mirror 
mass $m$. The noise spectral density reads~\cite{BK92}: 
\beq
\label{eq6}
S_{h}=\frac{1}{L^2}\,\left\{
       S_{{\cal Z} {\cal Z}}
      +2{\cal R}_{xx}\,\Re\left[S_{{\cal F} {\cal Z}}\right] 
+ {\cal R}_{xx}^2\, S_{{\cal F} {\cal F}}\right\}\,, 
\eeq
where the (one-sided) cross correlation of two operators is defined by  
$2\pi\delta\left(\Omega-\Omega'\right) S_{{\cal A} {\cal B}}(\Omega) =
\langle {\cal A}(\Omega) {\cal B}^\dagger(\Omega')+ {\cal B}^\dagger
(\Omega') {\cal A}(\Omega)\rangle$. 
Due to their dependence on $m$ the terms containing 
$S_{{\cal Z} {\cal Z}}$, $S_{{\cal F} 
{\cal F}}$ and $\Re\left[S_{{\cal F} {\cal Z}}\right]$ 
in Eq.~(\ref{eq6}) should be identified as the spectral densities of the 
effective shot noise, back-action noise and a term 
proportional to the effective correlation between the two noises \cite{BK92}.
{}From the definition of spectral density, one can derive 
\cite{BK92,BC200} the following uncertainty relation 
for the (one-sided) spectral densities and cross 
correlations  of ${\cal Z}$ and ${\cal F}$:
$S_{{\cal Z} {\cal Z}}\,S_{{\cal F} {\cal F}} - 
S_{{\cal Z} {\cal F}}\,S_{{\cal F} {\cal Z}} \ge \hbar^2$.
It turns out that this equation does not impose 
in general a lower bound on the noise spectral density (\ref{eq6}). 
However, in a very important type of interferometer
it does, namely for interferometers with uncorrelated shot noise and back-action noise, 
e.g., a LIGO-I/Virgo type conventional interferometer.
In this case 
$S_{{\cal Z} {\cal F}}^{\rm conv}=0=S_{{\cal F} {\cal Z}}^{\rm conv}$
\cite{KLMTV00} and inserting the vanishing correlations 
into Eq.~(\ref{eq6}) and into the uncertainty relation, one easily 
finds that $S_{h}^{\rm conv} \ge S_h^{\rm SQL}$. 
{}From this it follows that to beat the SQL one must 
build up correlations between shot noise and back-action noise.
In a SR interferometer the arm-cavity light containing the GW signal and 
the quantum-vacuum fluctuations enters the SR cavity through 
the dark port (see Fig.~\ref{Fig1}).
Part of this light leaks out through the SR mirror and contributes 
to the shot noise, but another portion, which is 
correlated to it, is fed back into the arm cavities 
and contributes to the radiation-pressure noise 
at some later time. This mechanism not only originates 
the nontrivial coupling between the antisymmetric mode of motion 
of the four arm-cavity mirrors and the signal-recycled 
optical field, which we discussed above, but also  
builds up dynamical correlations between the shot-noise and 
the radiation-pressure noise. Indeed, we obtain: 
$S_{{\cal Z} {\cal F}}^{\rm SR} =
S_{{\cal F} {\cal Z}}^{\rm SR} \neq 0$ (see Ref. \cite{BC200} 
for their explicit expressions).

In conclusion, our analysis has demonstrated
the importance of using fully quantum techniques to analyze
SR interferometers with LIGO-II parameters, where the correlations
between the shot and radiation-pressure noises are significant. It
also revealed the crucial role of the coupled optical-mechanical
dynamics in producing such correlations. 
It is now important to identify the best SR configuration, 
i.e. the choice of the physical parameters 
$I_0$, $\phi$, $\rho$, $\zeta$, and the readout scheme (homodyne or
modulation/demodulation) that optimizes the $S/N$
for various astrophysical GW sources. 

\vskip 0.5truecm
The authors thank K.S. Thorne for having introduced us to QND theory, 
for his constant encouragement and for very 
fruitful discussions and comments,  and V.B. Braginsky, 
F.~Ya. Khalili, Y. Levin and K.A. Strain for very 
helpful discussions and comments.
This research was supported in part by NSF grant PHY-9900776 and 
for AB also by the Richard C. Tolman Fellowship.

\end{document}